\documentclass[lettersize,journal]{IEEEtran}
\usepackage{algorithm2e}
\usepackage{amsmath,graphicx}
\usepackage{amssymb}
\usepackage{array}
\usepackage{bbm}
\usepackage{booktabs}
\usepackage{cite}
\usepackage[utf8]{inputenc}
\usepackage{mathtools}
\usepackage{mystyle}
\usepackage{multirow}
\usepackage{tikz}
\usepackage{hyperref}
\usepackage{pgfplots}
\usepackage{stfloats}
\usepackage[caption=false,font=normalsize,labelfont=sf,textfont=sf]{subfig}
\usepackage{textcomp}
\usepackage{tikz-3dplot}
\usepackage{listings,xcolor}
\usepackage{url}
\usepackage{verbatim}
\usetikzlibrary{arrows.meta}
\usetikzlibrary{positioning,through, backgrounds,decorations.pathreplacing,shapes.misc}
\pgfplotsset{compat=1.17}
\pgfplotsset{every axis/.append style={
                    label style={font=\Large},
                    tick label style={font=\Large}  
                    }}

\tdplotsetmaincoords{70}{30}

\definecolor{codegreen}{rgb}{0,0.6,0}
\definecolor{codegray}{rgb}{0.5,0.5,0.5}
\definecolor{codepurple}{rgb}{0.58,0,0.82}
\definecolor{backcolour}{rgb}{0.95,0.95,0.92}
\lstdefinestyle{lsty}{
    backgroundcolor=\color{backcolour},   
    commentstyle=\color{codegreen},
    keywordstyle=\color{magenta},
    numberstyle=\tiny\color{codegray},
    stringstyle=\color{codepurple},
    basicstyle=\ttfamily\footnotesize,
    breakatwhitespace=false,         
    breaklines=true,                 
    captionpos=b,                    
    keepspaces=true,                 
    numbers=left,                    
    numbersep=5pt,                  
    showspaces=false,                
    showstringspaces=false,
    showtabs=false,                  
    tabsize=1
}
\begin{document}

\title{End-to-end Open Vocabulary Keyword Search with~Multilingual Neural Representations}

\author{Bolaji Yusuf,~\IEEEmembership{Student~Member,~IEEE},~{Jan Černocký,~\IEEEmembership{Senior~Member,~IEEE}}~and~Murat Saraçlar,~\IEEEmembership{Member,~IEEE}
\thanks{This work was presented in part at the 22nd Annual Conference of the International Speech Communication Association, Brno, Czechia, September 2021~\cite{yusuf21_interspeech}.}
\thanks{Bolaji Yusuf is with Boğaziçi University Department of Electrical and Electronics Engineering, 34342 Istanbul, Turkey  (e-mail: bolaji.yusuf@boun.edu.tr) and also with Brno University of Technology, Faculty of Information Technology, Speech@FIT, 612 00 Brno, Czechia
}
\thanks{{Jan Černocký is with Brno University of Technology, Faculty of Information Technology, Speech@FIT, 612 00 Brno, Czechia (e-mail: cernocky@fit.vut.cz)}
}
\thanks{Murat Saraçlar is with Boğaziçi University Department of Electrical and Electronics Engineering, 34342 Istanbul, Turkey (e-mail: saraclar@boun.edu.tr)
}
}

\markboth{Journal of \LaTeX\ Class Files,~Vol.~14, No.~8, August~2021}%
{Shell \MakeLowercase{\textit{et al.}}: A Sample Article Using IEEEtran.cls for IEEE Journals}

\maketitle

\begin{abstract}
Conventional keyword search systems operate on automatic speech recognition (ASR) outputs, which causes them to have a complex indexing and search pipeline.
This has led to interest in ASR-free approaches to simplify the search procedure.
We recently proposed a neural ASR-free keyword search model which achieves competitive performance while maintaining an efficient and simplified pipeline, where queries and documents are encoded with a pair of recurrent neural network encoders and the encodings are combined with a dot-product.
In this paper, we extend this work with multilingual pretraining and detailed analysis of the model.
Our experiments show that the proposed multilingual training significantly improves the model performance and that despite not matching a strong ASR-based conventional keyword search system for short queries and queries comprising in-vocabulary words, the proposed model outperforms the ASR-based system for long queries and queries that do not appear in the training data.
\end{abstract}


\section{Introduction}
\label{sec:intro}
Keyword search (KWS) is one of the technologies that arose out of the need to efficiently index and search the ever-growing catalog of spoken content online.
Known alternatively as spoken term detection (STD), it entails locating short query phrases within large speech archives.
Given a written query, which may or may not have been encountered at training time, a KWS system is expected to return which utterances in the archive, if any, contain the query, the time stamps within those utterances hypothesized to correspond to the query and scores indicating the system's confidence in each hypothesis.

Conventional KWS involves using an automatic speech recognition (ASR) system to decode the archives, constructing an inverted index from the resulting lattices and searching the query therein~\cite{chelba2008retrieval}.
The inverted index is typically implemented as a timed factor finite-state transducer (FST)~\cite{can2011lattice,allauzen2004general}, which is constructed offline
and composed with an FST of the query.
While this approach has proven quite successful, operating downstream of ASR has its pitfalls.

One such pitfall is that indexing any utterance involves full ASR decoding, which incurs a nontrivial computational cost.
Furthermore, since the lattices of ASR systems can only contain tokens in the training vocabulary, ASR-based KWS systems with words as the ASR units cannot naturally retrieve out-of-vocabulary (OOV) queries, such as proper nouns and rare morphological inflections, and therefore have to resort to a host of other methods such as subword indexing or query expansion for retrieval~\cite{saraclar2004lattice,mamou2007vocabulary,parlak2012performance,szoke2008hybrid,chen2013using}.
Therefore, it is natural that recent research has focused on ASR-free KWS systems with a simpler indexing pipeline and natural handling of OOV queries~\cite{gundougdu2017joint,audhkhasi2017end}.
Since KWS is open-vocabulary and cannot therefore be cast as a keyword classification problem, these approaches typically feature a pair of encoders for speech and text trained to classify whether or not the text is spoken in speech segment under consideration.

In our recent conference paper~\cite{yusuf21_interspeech}, we proposed a dual-encoder-based keyword search model trained to predict frame-wise probabilities of existence of a query in an utterance.
It significantly outperformed other neural approaches in literature in terms of search accuracy, while also improving the search efficiency by using dot-products for search instead of more complicated feedforward neural networks.
This paper extends that preliminary work:
\begin{itemize}
    \item We conduct more comprehensive analysis of the model with experiments measuring the impacts of various components and parameters of the models as well as the performance of the models for various kinds of queries.
    \item We show that with slight modifications, the model can be trained multilingually and that finetuning such a multilingually-pretrained model significantly and consistently improves performance across target languages.
\end{itemize}

The rest of the paper is organized as follows:
Section~\ref{sec:related} covers previous related work and highlights their differences and, where appropriate, their similarities to our method;
Section~\ref{sec:methods} describes the proposed model;
Section~\ref{sec:experiments} details the experiments conducted and discusses the results of those experiments; Section~\ref{sec:conclusion} concludes the paper with a summary and future research directions.

\section{Related Work}
\label{sec:related}
\subsection{Open-vocabulary ASR-based KWS}
Since ASR language models are trained with a limited number of words due to computational and data availability constraints, the ASR output is limited to the closed vocabulary used in training.
However, user queries can---and often do---include words that are not part of this limited vocabulary.
Therefore, dealing with the challenge of seamlessly searching such OOV queries has been studied extensively within the context of ASR-based KWS systems.
The solutions in literature generally fall into two categories: using subword units and query expansion.

Based on the rationale that the words in a language---even OOV ones---can be composed from a limited set of subword units, using subword-based ASR has been the cornerstone of KWS in morphologically-rich languages~\cite{parlak2012performance,Turunen2008SpeechRF}, as well as open-vocabulary search in other languages~\cite{Ng2000,mamou2007vocabulary,karakos2014subword,su2015improvements}.
Most works use linguistic units such as syllables, morphs and phones, while others use data-driven units like graphones and multigrams~\cite{Turunen2008SpeechRF,arisoy2009turkish,he2016using}.
Although subword units increase the recall of the model, this comes at the cost of larger lattices which are more costly to index and search, as well as lower precision for in-vocabulary (IV) queries.
Therefore, it is common to use a hybrid of word lattices for IV queries and subword lattices for OOV ones.
This has the drawback of incurring the cost of double-indexing for every new utterance.
This drawback can be partly reduced by converting word lattices into subwords ones for OOV search~\cite{saraclar2004lattice,karakos2014normalization}, an approach limited in that it can only generate phone sequences which are substrings of some IV words.

Query expansion is an alternative approach to OOV search involving searching for phrases that are acoustically-similar to the query to account for ASR errors~\cite{can2009effect,chen2013using,saraclar2013confusion}
This is typically implemented by composing a query FST with an FST of phone confusions, before composing the expanded query FST with the index.
While query expansion can be used with subword indices, it can be leveraged to avoid double indexing by composing the expanded-query FST with the decoder vocabulary, resulting in acoustically-similar IV ``proxy" queries which can be searched in a word-based index.

While these approaches alleviate ASR-based KWS' inability to handle OOV queries, they invariably further complicate either the indexing or the search for the already complex ASR-based KWS system.
Our proposed method, on the other hand, not only offers simpler indexing and search than ASR-based KWS, it makes no distinction between IV and OOV queries.

\subsection{End-to-end Keyword Search}
Leveraging the ability of neural networks to model complex relationships, KWS traditionally comprising several disparate, separately-optimized, modules can now be simplified by formulating and optimizing appropriately designed neural architectures and objectives.
One such simplification involves using end-to-end ASR models to construct the KWS index as in~\cite{bai2016E2eAsrKws,audhkasi2018E2eAsrKws,shi2021timestamp,yang2022keyword}. 
While these works simplify ASR training, they still have complex KWS indexing pipelines since the simpler decoding algorithms for end-to-end ASR do not readily yield the timing and confidence information necessary for KWS.
Therefore, another direction, in which our work falls, involves training a model able to avoid the ASR decoding entirely while indexing or searching.

The authors of~\cite{gundougdu2017joint} propose a Siamese neural architecture which jointly learns a distance metric for speech documents represented as phone posteriorgrams along with a query representation and conduct search with subsequence dynamic time warping (DTW).
The method was extended in~\cite{gundogdu2018generative} to account for query dynamics and in~\cite{yusuf2019low} to learn better document representations.
While it showed impressive search accuracy, especially for OOV queries, this approach is limited in practice by the significant computational cost of DTW.

The authors of~\cite{Sacchi2019} similarly proposed a Siamese architecture which learns text and speech representations for the related task of open-vocabulary hotword spotting.\footnote{Hotword spotting involves spotting a limited set of phrases and has also been referred to as keyword spotting in some literature. We avoid that term to avoid confusion as some other literature use keyword spotting to refer to keyword search of the kind that we tackle.}
Since the task there does not involve localization of the keywords, there is no associated cost of DTW.
In~\cite{bluche20_interspeech}, a meta-network was proposed that, for a given query, generates the parameters of a model to classify whether or not a speech segment contains that query.
Thus the parameters of the model grow with the number of queries, which makes the model more suitable for limited, but adaptable, query sets as opposed to the unlimited vocabulary as in KWS.
Moreover, like the model featured in ~\cite{Sacchi2019}, it also lacks the ability to localize the queries, which makes both approaches unsuitable for keyword search where the timestamps of each query's occurrence are required.

In~\cite{audhkhasi2017end}, a model was proposed with a pair of encoders for computing fixed-length representations of speech utterances and text respectively, and a feedforward search network classifying whether or not the encoded text occurs in the encoded utterance.
While the model was innovative in showing that the open-vocabulary search pipeline can be greatly simplified with this dual-encoder structure, it was limited to the utterance classification task where the probability of existence of a query was artificially set to 0.5 (by sampling positive and negative test utterances with equal probability at test time) and could not work in the highly imbalanced scenario of realistic KWS, where the number of negative trials far outnumber the number of positive ones.
Moreover, since the speech encoder outputted a fixed-length representation of each utterance, the model could not temporally localize the keywords, although the authors did experiment with a coarse form of localization by classifying whether the query occurs in the first or second half of the utterance.
This method was improved in~\cite{zhao2020end} by using better pretraining objectives, although it still had the limitations of~\cite{audhkhasi2017end} with regards to handling realistic KWS settings.

The most relevant related work to ours are those in~\cite{fuchs2021cnn} and~\cite{svec21_interspeech} who contemporaneously with us proposed dual encoder architectures capable of open-vocabulary keyword search in realistic settings, complete with the ability to temporally localize the queries.
Like us, they ensure that the speech encoders do not lose temporal information, and use forced alignment to obtain the query locations at training time.

In~\cite{fuchs2021cnn}, the authors extend their prior work on closed-set keyword detection and localization~\cite{Segal2019} to cover open vocabularies.
A feedforward network takes the speech and text encodings from a pair of convolutional and recurrent encoders and returns a vector to be compared with the text encoding to determine whether that text occurs in the utterance, as well as a pair of floating points corresponding to the hypothesized locations of the query.
By directly predicting the temporal locations, they avoid the need to have any post processing step.
This however comes at the cost of having to run the feedforward network for every query-utterance pair at search time, as opposed to the simpler matrix-vector product that we use for the query-utterance interaction.
Therefore, the search can become orders of magnitude slower since each feedforward layer with output dimension of $F$ is a matrix-matrix product having $F$ times the cost of matrix-vector product.

In~\cite{svec21_interspeech}, the authors propose a very similar structure to our model.
The main difference is that, instead of acoustic features, they use phonetic confusion networks output from an external ASR system as the speech representation.
This allows language model information to be indirectly incorporated into the KWS model.
However, it also means that creating the document representation requires full ASR decoding.
Note that while we also use bottleneck features (BNF) obtained from an external ASR model, extracting BNF only incurs the cost of passing data through the acoustic model and not the costs of searching the ASR decoding graph.

\subsection{Multilingual data for KWS}
Using multilingual data to improve KWS has been explored in prior work in the context of both ASR-based and ASR-free KWS.
In the context of ASR-based KWS systems, a common recipe is to improve the ASR, and consequently KWS, in low-resource settings by training the acoustic model multilingually~\cite{knill2013investigation,tuske2014data,cui2015multilingual,Karafiat2017,sercu2017network}.
This generally entails sharing the lower layers of the acoustic model and using either a shared output layer~\cite{thomas2010cross,Vu2012,vu2014multilingual} or separate output layers for each language~\cite{scanzio2008use,Grezl2011,Vesely2012,heigold2013multilingual}.

In~\cite{yusuf2019low}, a joint metric and representation learning method is used to incorporate multilingual bottleneck features into dynamic-time-warping-based KWS.
Multilingual bottleneck features and posteriorgrams have also been used for query-by-example (where both query and audio are spoken) with~\cite{rodriguez2014high,szoke2015coping,chen2017multitask} or without~\cite{ram2020neural} dynamic time warping.
While these works use multilingual data to train the feature extractor for ASR-free search, they do not train the search model itself multilingually.

\section{Methods}
\label{sec:methods}
\begin{figure}[t]
    \centering
    \includegraphics[width=\linewidth]{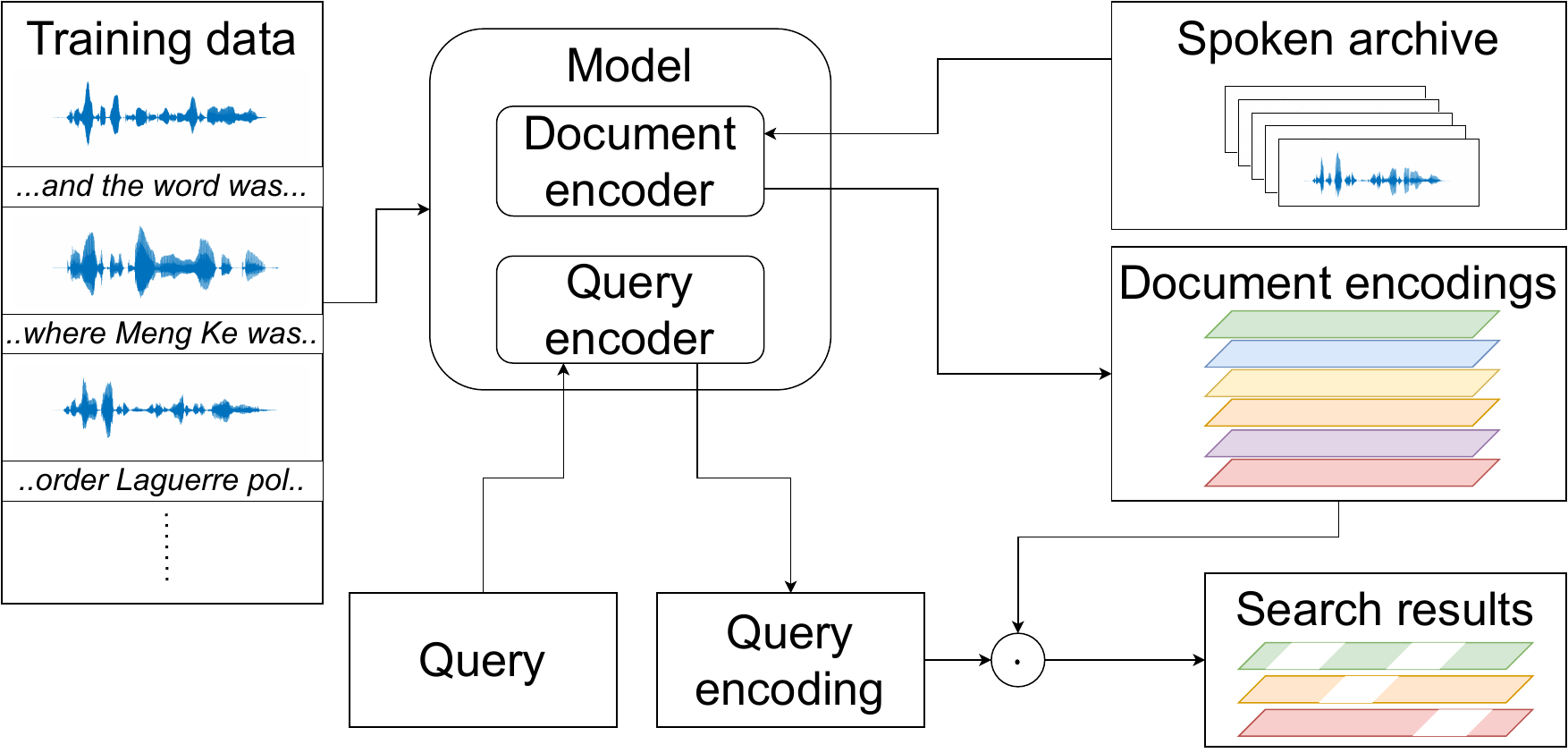}
    \caption{
    Flowchart illustrating the proposed system.
    A keyword search model, comprising a query encoder and a document encoder, is trained on forced-aligned speech and text data.
    The document encoder is used to encode spoken archives for efficient search, and
    the query encoder converts each query into a vector form which is searched by computing frame-wise inner products with the document representation.}
    \label{fig:flowchart}
\end{figure}

In this section, we describe the method we have proposed for keyword search.
The method, illustrated in Figure~\ref{fig:flowchart}, involves a soft-indexing and matching approach.
Where ASR-based KWS methods index a spoken archive by decoding it into a graph of symbolic units and conduct search by matching with a corresponding graph of the query, our approach uses a dense representation in a vector space and conducts search by matching the query with dot-products, for which modern CPUs, GPUs and linear algebra libraries have efficient implementations.

Sections~\ref{subsec:forumulation}~to~\ref{subsec:postprocess} replicate the model definition, training and search procedures from~\cite{yusuf21_interspeech} for completeness and readers' convenience.
Section~\ref{subsec:multilingual_training} describes multilingual training.

\subsection{Problem formulation}
\label{subsec:forumulation}
We formulate keyword search as the task of classifying whether a keyword occurs at any given location in the document.
Given a query phrase $\Matrix{q} = \bigl(q_1, q_2, \dots, q_K \bigr)$ where each $q_k$ is a letter, and an utterance represented as a sequence of acoustic features $\Matrix{X} = \bigl(\Matrix{x}_1, \dots, \Matrix{x}_{N_x}\bigr)$, we seek the sequence of occurrence indicators $\Matrix{y}{(\Matrix{q}, \Matrix{X})}= (y_1, \dots, y_{N_x}) \in \{0, 1\}^{N_x}$ such that:
\begin{align}
    y_n=
    \begin{cases}
      1, & \text{if}\ \Matrix{q} \text{ is spoken in } \Matrix{X} \text{ in a time span including } n \\
      0, & \text{otherwise}.
    \end{cases}
\end{align}
Given a training set of utterances $\mathcal{X} = \{\Matrix{X}^{(1)}, \Matrix{X}^{(2)}, \dots, \Matrix{X}^{(S)}\}$ and query phrases $\mathcal{Q} = \{ \Matrix{q}^{(1)}, \Matrix{q}^{(2)}, \dots, \Matrix{q}^{(L)} \}$,
we train a neural network with parameters $\Matrix{\theta}$ to minimize the negative log-likelihood of the occurrence indicators:
\begin{align}
    \Matrix{\theta}^* = \argmin_{\Matrix{\theta}} \sum_{ l=1 }^L \sum_{ s = 1 }^S \sum_{ n = 1}^{N_{x{(s)}}} -\log p_{\Matrix{\theta}} \Bigl( y_n | \Matrix{q}^{(l)}, \Matrix{X} ^ {(s)} \Bigr).
    \label{eq:posterior}
\end{align}
Note that training requires timing information of the phrases in the training set. We obtain it by training an HMM-GMM-based ASR system on the training data and using it to generate the required word-level forced alignments.

\subsection{Model definition}

\begin{figure}[t]
    \centering
    \includegraphics[width=\linewidth]{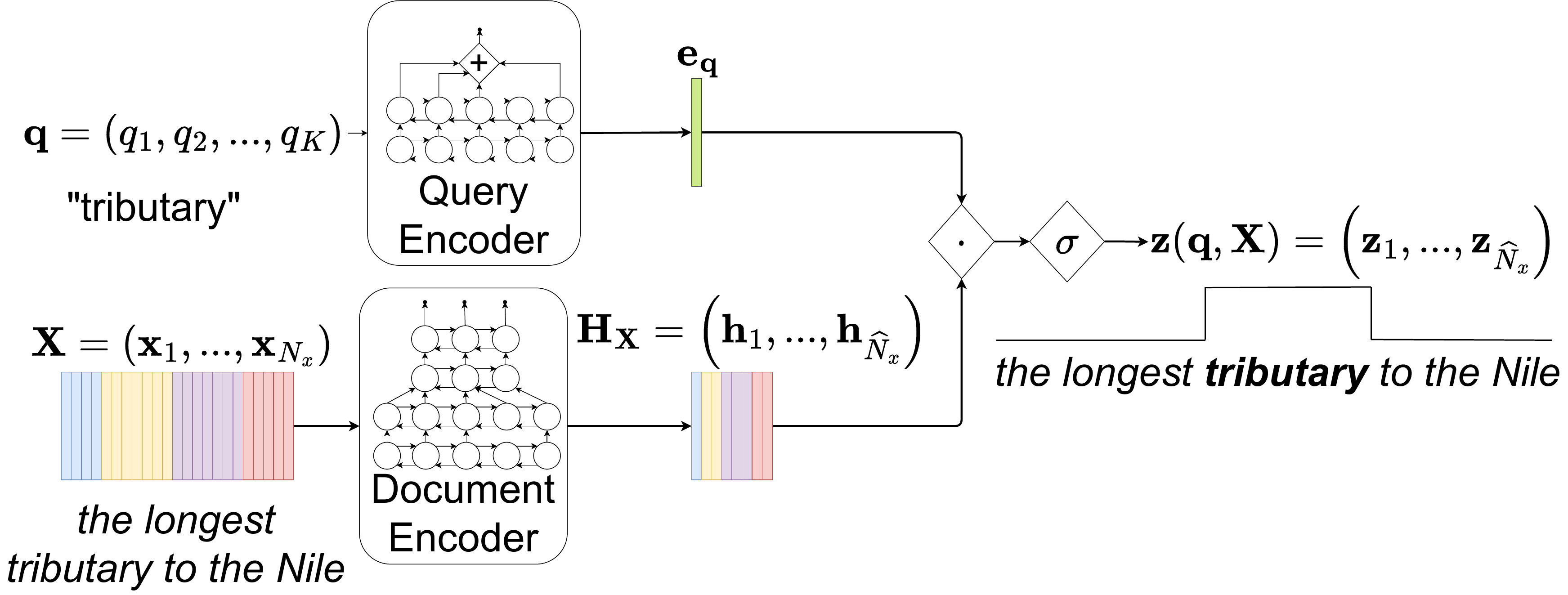}
    \caption{
    Overview of the neural keyword search model.
    The document encoder is a bidirectional recurrent neural network which outputs a subsampled representation of the spoken document.
    The query encoder is a recurrent neural network which outputs a fixed length representation of the written query.
    A dot-product between the two representations gives logits which are passed through a logistic sigmoid to predict the likelihood of the query occurring at each subsampled time frame.
    }
    \label{fig:model}
\end{figure}

Our keyword search model, depicted in Figure~\ref{fig:model}, comprises a recurrent document encoder and a recurrent query encoder.
We conduct the search via a matrix-vector multiplication between the outputs of the encoders.
We apply the logistic sigmoid function to the resulting vector of logits to obtain frame-wise posterior probabilities $p_{\Matrix{\theta}}(y_n|\dots)$ which we post-process to detect the locations of each keyword.
Since the document encoder is intended to act as an offline indexer, while the query encoder must be called whenever a query is received, we ensure that the query encoder is a much smaller neural network than the document encoder.

\subsubsection{Query encoder}
The input to the query encoder is a sequence of letters $\Matrix{q} = (q_1, \dots, q_K)$ that constitute the query and the output is a fixed-length representation $\Matrix{e}_{\Matrix{q}} \in \mathbb{R}^D$.
A trainable input embedding layer converts the sequence of letters into a sequence of vectors that are input into a stack of bidirectional gated recurrent unit (GRU) layers.
The GRU outputs another sequence of vectors $\Matrix{V} = (\Matrix{v}_1, \dots, \Matrix{v}_K)$.
The final query representation is then computed from the sum of these GRU output vectors along the sequence axis:
\begin{align}
    \Matrix{e}_{\Matrix{q}} = \sum_{k=1}^{K} \Matrix{W}_1 \Matrix{v}_k + \Matrix{b_1},
    \label{eq:query_rep}
\end{align}
where $\Matrix{W}_1$ and $\Matrix{b}_1$ are the weight and bias of a trainable affine transform that changes the dimensionality of the query representation to ensure it matches the output of the document encoder.
The affine transform also ensures that the dynamic range of the query encoding is not limited by the hyperbolic tangent output of the GRU to $(-1, 1)$.

We use summation instead of taking the output of the GRU at the final step, i.e., instead of setting $\Matrix{e}_{\Matrix{q}} = \Matrix{W}_1 \Matrix{v}_K + \Matrix{b_1}$, because we empirically found it to be better.
We also experimented with having a unidirectional query encoder but we found the bidirectional encoder to be superior.

\subsubsection{Document encoder}
The input to the document encoder is the sequence of speech features $\Matrix{X}$ of length $N_x$.
First, $\Matrix{X}$ is passed through a stack of bidirectional long short term memory (BLSTM) layers which output $\Matrix{U} = (\Matrix{u}_1, \dots, \Matrix{u}_{\hat{N}_x})$ of length $\hat{N}_x$.
We down-sample the hidden representations between some of the BLSTM layers so that $\hat{N} = \floor{\frac{N}{4}}$.
This decreases the computational cost of storage and search, and we found empirically that it improves the search accuracy as it reduces the durations processed by higher layers, making it easier to model long-range dependencies.
The final encoder output is then $\Matrix{H}_{\Matrix{X}} = (\Matrix{h}_1, \dots, \Matrix{h}_{\hat{N}_x})$, such that for each $\hat{n}$:
\begin{align}
    \Matrix{h}_{\hat{n}} = \Matrix{W}_2 \Matrix{u}_{\hat{n}} + \Matrix{b}_2,
\end{align}
where $\Matrix{W}_2$ and $\Matrix{b}_2$ constitute an affine transformation similar to that at the output of the query encoder.

We choose to make the query encoder a GRU instead of an LSTM in order to reduce its computational cost since, unlike the document encoder which we expect to operate as an offline indexer, the query encoder would be run whenever a user queries the system.
Moreover, in preliminary experiments, we found using a GRU as the query encoder performs as well as an LSTM with the same dimensions while having three-quarter the size and computational cost of the latter.
However, using a GRU as the document encoder instead of an LSTM significantly degrades search performance.

\subsubsection{Search function}
The search output is given by multiplying the encoded document matrix with the encoded query vector, followed by a logistic sigmoid, resulting in the desired vector of per-frame occurrence probabilities $\Matrix{z}(\Matrix{q}, \Matrix{X})=(z_1, \dots, z_{\hat{N}}) \in (0, 1)^{\hat{N}}$, where $z_n \coloneqq p_{\Matrix{\theta}} \bigl( y_n | \Matrix{q}, \Matrix{X} \bigr)$:
\begin{align}
    \Matrix{z}(\Matrix{q}, \Matrix{X}) = \sigma(\Matrix{H}_{\Matrix{X}} ^\top \Matrix{e}_{\Matrix{q}}).
    \label{eq:output}
\end{align}
Since the document and query representations only interact through this product and are otherwise independent, we can pre-compute and store the encodings of the documents.
Thus, at search time, only the cost of computing the query representation (from the much smaller query encoder) and the cost of the dot product is incurred.

As the only interaction between the query and a time frame of the document is this inner-product, which is independent of other time frames at the encoding level, the document encoding at each frame clearly needs to encode enough information to disambiguate between similar queries.
For example, if the document contains the word ``predict", the encodings of each frame corresponding to ``pre-" need to be distinguishable from the encodings of ``pre-" in, say, ``prelude" or ``preface".
Similarly, the encodings corresponding to ``-sion" in ``confusion", need to be distinguishable from those in ``television" and ``intrusion".
Therefore the document encoder needs to be bidirectional because the LSTM's forward direction is needed to disambiguate between shared suffixes, while the reverse direction is necessary for separating phrases with shared prefixes.
In other words, the reverse direction controls when occurrence probabilities should start spiking and the forward direction controls when they should stop spiking.

\subsection{Model training}
\label{sec:methods:training}
Training the model involves optimizing~\eqref{eq:posterior} with gradient descent.
However, doing so directly is impractical as computing the gradient involves summation over all phrases and utterances in the training set.
A corpus of ${S}$ utterances with ${W}$ words each has $\mathcal{O}({S} {W}^2)$ elements in the double summation.
Therefore, we approximate this large sum with a smaller sum whose gradient approximates the gradient of the original and optimize the approximate sum instead:
\begin{align}
    \Matrix{\theta}^* \approx \argmin_{\Matrix{\theta}} \sum_{ l=1 }^{L_b} \sum_{ m = 1 }^M \sum_{ n = 1}^{N_{x{(m)}}} -\log p_{\Matrix{\theta}} \Bigl( y_n | \Matrix{q}^{(l)}, \Matrix{X} ^ {(m)} \Bigr),
    \label{eq:posterior_approx}
\end{align}
where $L_b \ll L$ is the mini-batch size for each training step, and
$M = |\mathcal{X}_{\Matrix{q}^{(l)}}| \ll S$ is the number of utterances sampled for each training phrase.

When looping over the phrases in the training data, we only consider such $l$s that $\Matrix{q}^{(l)}$ is either a unigram, bigram or trigram.
When sampling the utterances $\mathcal{X}_{\Matrix{q}^{(l)}}$ for each step, we ensure that at least one of them is a ``positive" utterance, i.e., it contains the training phrase $\Matrix{q}^{(l)}$, while the others are sampled truly randomly.
While this constraint biases the gradient, without it, an overwhelming majority of mini-batches would be ``negative", i.e., have all-zero labels, which would make optimization impossible.

For each query-utterance training pair $(\Matrix{q}^{(l)}, \Matrix{X}^{(m)})$, we minimize an objective function $J(\Matrix{q}^{(l)}, \Matrix{X}^{(m)})$ between the sigmoid outputs $\Matrix{z}(\Matrix{q}^{(l)}, \Matrix{X}^{(m)})$ and the labels $\Matrix{y}(\Matrix{q}^{(l)}, \Matrix{X}^{(m)})$:
\begin{align}
    J \Bigl( \Matrix{q}^{(l)}, \Matrix{X}^{(m)} \Bigr) = -\sum_{n=1}^{\hat{N}_{x{(m)}}}& \Bigl (\mathbbm{1}_{z_n > 1 - \phi} \cdot (1-y_{n}) \log (1-z_n)
     \nonumber \\
    & + \mathbbm{1}_{z_n < \phi} \cdot \lambda \cdot y_{n} \log z_n \Bigr),
    \label{eq:loss_single}
\end{align}
where the labels have been down-sampled to match the output frame rate of the document encoder.
This objective function extends the binary cross-entropy objective with the hyper-parameters $\lambda$ and $\phi$.
When both are set to $1$, the loss reduces to the binary cross-entropy.
$\lambda$ controls the relative importance of frames labeled $1$ and frames labeled $0$, i.e., the relative cost of misses to false detections; as $\lambda$ increases, frames labeled $1$ contribute more to the total loss.
$\phi$ is a strictness term controlling the sensitivity of the loss function to easily classified frames; frames labeled $1$ with sigmoid outputs above $\phi$ and frames labeled $0$ with sigmoid outputs below $1-\phi$ do not contribute to the loss.
This prevents the model from learning to better classify frames that are already well classified at the expense of learning to classify difficult frames.

\subsection{Post-processing for keyword search}
\label{subsec:postprocess}
Having obtained the vector of probabilities from~\eqref{eq:output}, we still need to post-process them to obtain the timestamps in the document hypothesized to contain the query, and the corresponding confidence scores.
The procedure is as follows:
\begin{enumerate}
    \item We zero-out the probabilities ($z_n$) below some threshold $\alpha$.
    This thresholding is a necessary first step because it is otherwise impossible to determine discrete values, $\{0, 1\}$, of $y_n$ since sigmoid outputs are strictly non-zero.
    We treat $\alpha$ as a hyper-parameter which we tune on development sets to select from among $\{0.2, 0.4, 0.6\}$.
    \item We pick the resulting ``islands" of non-zero elements as our system hypotheses.
    Each hypothesis' confidence score is computed as the median probability in its interval.
    We also experimented with the mean and max operations but found median to be better.
\end{enumerate}

\subsection{Multilingual training}
\label{subsec:multilingual_training}
Neural approaches, in KWS and otherwise, generally struggle in low-resource settings since they require large amounts of data to train.
We therefore explore multilingual pretraining to improve the performance of the proposed model in low-resource settings.
To do this, we pretrain the model with data pooled from several letter-based languages, and then finetune it on the target language.

During multilingual training, the entire model is shared by all languages, with the exception of the query encoder's input embedding layer.
The model is trained with the same objective as in the monolingual setting.
The negative utterances sampled for each training phrase can come from any language; we experimented with enforcing that the negative utterances come from the same language but found that this did not lead to consistent improvements.

When finetuning to a new language, we transfer only the pretrained document encoder and reinitialize the entire query encoder with random weights, then train the entire model on the target language's training data.
We experimented with transferring the multilingually-pretrained query encoder as well with only the embedding layer reinitialized, but found this to result in significantly worse performance.
We also experimented with initially freezing the transferred document encoder for a few epochs so that the query encoder gets trained to a reasonable degree before finetuning the whole model, but we found doing so also worsened performance and made training unstable.

\section{Experiments}
\label{sec:experiments}
In this section, we conduct experiments to analyze various components of the proposed model.
First, we describe experiment setup: the datasets, metrics and default hyper-parameters for the proposed model.
Then we analyze how various parameters affect the model performance.
Finally, we analyze how the performance of the model changes with different keyword properties and compare and contrast to how those same keyword properties affect a conventional LVCSR-based model.

\subsection{Experimental setup}
\subsubsection{Dataset}
\label{sec:experiments:dataset}
We conduct all experiments on data from the IARPA Babel corpus.\footnote{https://www.iarpa.gov/index.php/research-programs/babel}
We use the limited language packs of Pashto, Turkish and Zulu for KWS experiments.
Each of these comprises 10 hours of transcribed data for training, a 10-hour development set for hyper-parameter tuning and a 5-hour evaluation set.\footnote{The full evaluation set used for Babel challenges is 15 hours, of which the references for only 5 hours are openly available.}
Table~\ref{tab:iv_oov_dist} shows the distribution of queries for each language's development and evaluation sets.
We use the transcribed data from 16 other Babel languages---totaling 170 hours---for multilingual pretraining of the KWS model in~\ref{sec:experiments:performance}.
We also use this dataset to train a multilingual acoustic model which contains a bottleneck layer for extracting 42-dimensional bottleneck features.
The BNF extractor has language-specific output layers trained to predict the pretraining languages' senone labels, where each language's context dependent triphones clustered to around 2000 such senones.
\begin{table}[t]
    \centering
    \caption{Distribution of in-vocabulary and out-of-vocabulary queries in each test language.}
    \begin{tabular}{lcccccccc}
    \toprule
    Query type & \multicolumn{2}{c}{Pashto} && \multicolumn{2}{c}{Turkish} && \multicolumn{2}{c}{Zulu} \\
    & Eval & Dev && Eval & Dev && Eval & Dev \\
    \midrule
         IV &  2044 & 1319 && 1173 & 203 && 1028 & 1193 \\
         OOV & 326 & 438 && 452 & 80 && 380 & 793 \\
         \bottomrule
    \end{tabular}
    \label{tab:iv_oov_dist}
\end{table}

\subsubsection{Metrics}
We report results in terms of the variants of the term weighted value (TWV)~\cite{Fiscus2006}.
The actual term weighted value (ATWV) is a measure of weighted precision and recall at a single predefined threshold. For a set of queries $\mathcal{Q}$ and threshold $\xi$, the ATWV is defined:
\begin{equation}
ATWV(\xi, \mathcal{Q}) = 1-\frac{1}{|\mathcal{Q}|}\sum_{q \in \mathcal{Q}}(P_{miss}(q,\xi) + \beta P_{FA}(q,\xi)) ,
\label{eqn:twv}
\end{equation}
where $P_{miss}(q,\xi)$ is the probability of misses, $P_{fa}(q,\xi))$ is the probability false alarms, and $\beta$ is a parameter that controls the relative weights of false alarms and misses.
Following the NIST STD evaluations~\cite{Fiscus2007}, we set $\beta=999.9$.
On the development set, we report the maximum term weighted value (MTWV) which is the TWV at the threshold that maximizes it.
This threshold is then used to compute the actual term weighted value (ATWV) for the evaluation set.

Since different queries tend to have different score distributions and term weighted value requires setting a single global threshold, it is necessary to normalize scores per query.
To this end, we adopt the keyword specific thresholding normalization method from~\cite{miller2007rapid}.

We also report the optimum term weighted value (OTWV)---the upper-bound MTWV computed with query-specific thresholds.
OTWV gives a measure of term weighted value without the effects of inter-query score mis-calibration.

Finally, we measure the supremum term weighted value (STWV)---the OTWV with the cost of false alarms set to zero.
This gives a measure of overall recall.
We however limit our use of STWV except when comparing two systems with similar ATWV because without such a constraint, STWV can be inflated by simply ``detecting" the query everywhere.

Term weighted values so defined have a theoretical maximum value of $1$.
In our results, we multiply all term weighted values by $100$ to get scores that can go up to $100$.

\subsubsection{Model configuration and default hyper-parameters}
In general, we base our default architecture off that described in~\cite{yusuf21_interspeech}.
The document encoder has 6 BLSTM layers with 512 output units, followed by an affine projection layer with output dimension of 400.
We apply dropout of 0.4 between successive BLSTM layers and down-sample by a factor of 2 after the first and fourth BLSTM layers.
The query encoder has a 32-dimensional input embedding layer, 2 bidirectional GRU layers with 256 units each and a projection layer to match the output of the document encoder.
This is almost identical to the setup in~\cite{yusuf21_interspeech} except that we remove all Batchnorm layers, as we found finetuning multilingual models trained with Batchnorm to be unstable.
Moreover, in the monolingual setting, we did not notice any performance deterioration from excising the Batchnorm layers.

We use graphemes as the query input representations instead of phonemes as they remove the need to train any grapheme-to-phoneme (G2P) converters to use for OOV queries.
Moreover, we found them empirically superior to phonemes in terms of KWS performance.
We do however concede that different representations would be required for languages that have orthographies with no phonetic correspondence.

Except where stated otherwise, we set $\lambda=5$ and $\phi=0.7$ in Equation~\ref{eq:loss_single} and use 3 negative examples per positive at training time, i.e., $M=4$ in Section~\ref{sec:methods:training}.

\subsection{Effect of loss function parameters}
\label{sec:experiments:loss}
\begin{figure}
    \hspace{-0.12\linewidth}
    \includegraphics[width=1.2\linewidth]{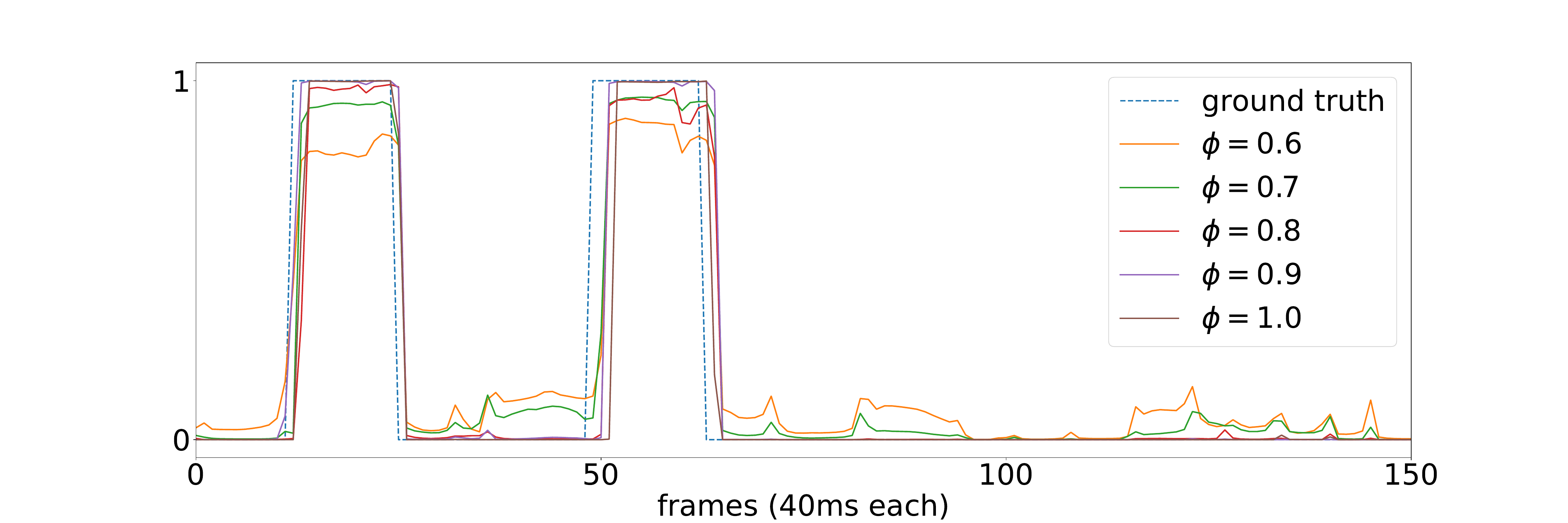}
    \caption{Outputs from the KWS model for an example query-utterance pair from the Turkish development set under various settings of the training objective tolerance ($\phi$).}
    \label{fig:model_output_examples}
\end{figure}
In this section, we analyze the impact of the hyper-parameters of the loss function, $\phi$ and $\lambda$ from Equation~\ref{eq:loss_single}.
Remember that $\lambda$ is the weight given to positive training frames (frames which contain the training phrase), while $\phi$ controls the allowable margin beyond which no loss is incurred.
When both values are set to 1, the objective becomes the classic binary cross-entropy objective.
All models here use bottleneck features as input without any pretraining or speed perturbation.

First we vary the tolerance ($\phi$) of the loss function with $\lambda$ fixed to $5$.
Figure~\ref{fig:model_output_examples} depicts the output of models trained with various values of $\phi$ on an example from the Turkish dev set.
All settings of $\phi$ track the correct shape, outputting high values where the ground truth is $1$ and low values where the ground truth is $0$.
However, the degrees with which they do so vary, with increasing $\phi$ expectedly resulting in more extreme separation of positives from negatives.

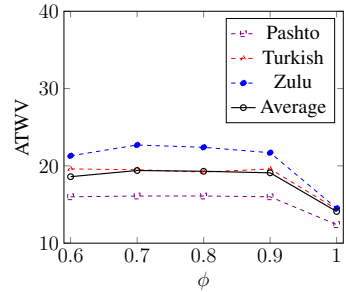
\begin{figure}[t]
    \centering
    \resizebox{0.25\textwidth}{!}{%
        \begin{tikzpicture}
            \begin{axis}[
    xlabel={$\phi$},
    ylabel={ATWV},
    xmin=0.59, xmax=1.01,
    ymin=10.0, ymax=40,
    legend pos=north east,
    legend style={nodes={scale=1.4, transform shape}}
]

    \addplot[color=violet, dashed, mark=square,]
    coordinates {
        (0.6, 16.0) 
        (0.7, 16.1)
        (0.8, 16.1)
        (0.9, 16.0)
        (1.0, 12.4)
    };
    \addlegendentry{Pashto}

    \addplot[color=red, dashed, mark=triangle,]
    coordinates {
        (0.6, 19.6) 
        (0.7, 19.5)
        (0.8, 19.2)
        (0.9, 19.6)
        (1.0, 14.4)
    };
    \addlegendentry{Turkish}

    \addplot[color=blue, dashed, mark=*,]
    coordinates {
        (0.6, 21.3) 
        (0.7, 22.7)
        (0.8, 22.4)
        (0.9, 21.7)
        (1.0, 14.5)
    };
    \addlegendentry{Zulu}

    \addplot[color=black, mark=o, thick]
    coordinates {
        (0.6, 18.6) 
        (0.7, 19.4)
        (0.8, 19.3)
        (0.9, 19.1)
        (1.0, 14.1)
    };
    \addlegendentry{Average}
\end{axis}
        \end{tikzpicture}
    }
    \caption{ATWV on the evaluation sets as the strictness term $\phi$ in the training objective is varied.}
    \label{fig:margin}
\end{figure}

Figure~\ref{fig:margin} shows the ATWV as $\phi$ changes.
We observe that ATWV does not vary much with the choice of $\phi$ except when it is set to $1$ where we observe significant ATWV degradation.
This implies that the exact value of $\phi$ is not as important so long as we have some tolerance.
Although we do not report those results here, we found that setting higher values of $\phi$ allows us to use lower sigmoid thresholds than $0.5$, resulting in higher STWV without sacrificing ATWV.

\begin{figure}[t]
    \centering
    \resizebox{0.25\textwidth}{!}{%
        \begin{tikzpicture}
            \input{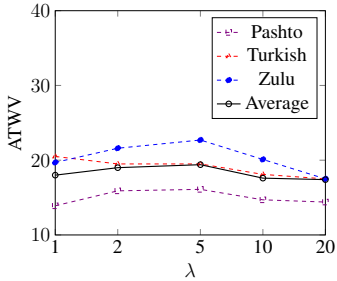}
        \end{tikzpicture}
    }
    \caption{ATWV on the evaluation sets as the weight of positive frames $\lambda$ in the training objective is varied.}
    \label{fig:loss_weight}
\end{figure}

Figure~\ref{fig:loss_weight} shows the ATWV as $\lambda$ is varied with $\phi$ fixed to $0.7$.
The ATWV increases with $\lambda$ until around $5$ where it peaks, and then falls off as $\lambda$ is further increased.
This indicates that the decrease in precision that accompanies increasing $\lambda$ starts to hurt the overall ATWV.
Thus, the correct setting of $\lambda$ seems tied to the relative costs of false alarms to misses and must be set based on the task at hand.
However, we note the recall, as measured by STWV---which is not in the figure---increases monotonically with $\lambda$, which is expected as increasing $\lambda$ makes the training objective prioritize detecting positive locations over suppressing negative ones.

\subsection{Effect of down-sampling}
\label{sec:experiments:sampling}
Our document encoder features a pair of down-sampling steps after the first and fourth BLSTM layers.
Thus, the output document encodings are down-sampled by a factor of $4$, resulting in smaller document storage and computational requirements.
In this section, we analyze the effect of down-sampling on keyword search performance.

\begin{figure}[t]
    \centering
    \resizebox{0.25\textwidth}{!}{%
        \begin{tikzpicture}
            \input{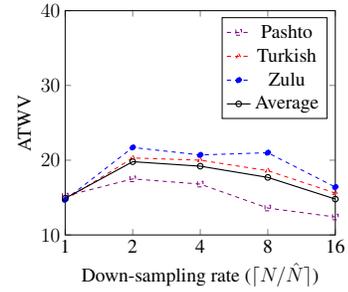}
        \end{tikzpicture}
    }
    \caption{ATWV on the evaluation sets as the down-sampling rate is varied.}
    \label{fig:subs}
\end{figure}

Figure~\ref{fig:subs} shows the ATWV as the total down-sampling factor varies between 1 (-),\ 2 (4),\ 4 (1, 4),\ 8 (1, 3, 4) and 16 (1, 3, 4, 5) where the numbers in the parentheses denote which layers' outputs are down-sampled by $2$.
We note that the ATWV improves as we introduce down-sampling, decreases slightly as the down-sampling factor is increased from $2$ to $4$, and starts to degrade upon further down-sampling---with a down-sampling factor of $16$, the ATWV gets worse than not having any down-sampling at all.

Overall, we infer that having down-sampling---within some bounds---does not just maintain accuracy but actually improves it, while being faster.
We ascribe this improvement to the difficulty of learning very long range dependencies within the model encodings.
For instance, to correctly localize a query of length 500ms at a frame rate of 10ms, each BLSTM hidden state would have to contain information spanning at least 50 frames; after down-sampling by a factor of 2, this burden reduces to 25 frames etc.
Beyond the optimal down-sampling rate, the search fidelity starts to degrade, ostensibly due to the excessive loss in resolution.

\begin{figure}[t]
    \centering
    \resizebox{0.4\textwidth}{!}{%
        \begin{tikzpicture}
            \centering
  \begin{axis}[
        ybar, axis on top,
        height=8cm, width=15.5cm,
        bar width=0.4cm,
        ymajorgrids, tick align=inside,
        major grid style={draw=white},
        enlarge y limits={value=.1,upper},
        axis x line*=bottom,
        xlabel={Query length in letters},
        y axis line style={opacity=0},
        tickwidth=0pt,
        enlarge x limits=true,
        legend style={
            at={(0.35,0.9)},
            anchor=north,
            legend columns=-1,
            /tikz/every even column/.append style={column sep=.5cm}
        },
        legend style={nodes={scale=1.5, transform shape}},
        ylabel={ATWV difference from DS=1},
        symbolic x coords={
           $\le 4$,5-7,8-10,11-13,14-16,$\ge 17$},
       xtick=data,
    ]

    \addplot [draw=none, fill=blue!30] coordinates {
      ($\le 4$, 0.2688888888888889)
      (5-7, 3.1476388888888884)
      (8-10, 5.619305555555556)
      (11-13, 4.463819444444443)
      (14-16, 11.327847222222223)
      ($\ge 17$, 12.124166666666667) };
   \addplot [draw=none,fill=red!30] coordinates {
      ($\le 4$, -0.1345833333333333)
      (5-7, 2.442013888888889)
      (8-10, 5.347638888888889)
      (11-13, 5.28833333333333)
      (14-16, 9.386111111111111)
      ($\ge 17$, 10.490833333333335) };
   \addplot [draw=none, fill=green!30] coordinates {
      ($\le 4$, -0.730787037037037)
      (5-7, 0.7550694444444449)
      (8-10, 3.036527777777778)
      (11-13, 6.325624999999999)
      (14-16, 12.235972222222225)
      ($\ge 17$, 11.458958333333335) };
    \addplot [draw=none, fill=black!30] coordinates {
      ($\le 4$, -0.7193055555555555)
      (5-7, -0.0800694444444442)
      (8-10, -0.385555555555555)
      (11-13, -1.3995138888888905)
      (14-16, 6.462013888888887)
      ($\ge 17$, 5.441666666666666) };

    \legend{DS=2,DS=4,DS=8,DS=16}
  \end{axis}
        \end{tikzpicture}
    }
    \caption{Average difference in ATWV of systems with various down-sampling factors when compared to the system with no down-sampling. DS=* denotes down-sampling factor of *.
    }
    \label{fig:atwv_kwlen_down_1_2_4}
\end{figure}
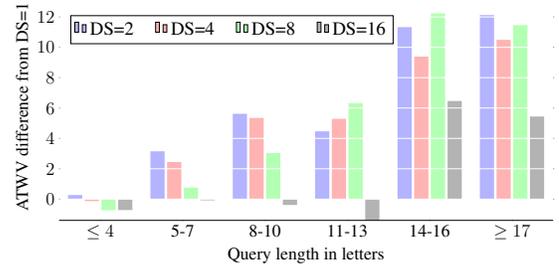

These arguments are supported by the results in Figure~\ref{fig:atwv_kwlen_down_1_2_4} which shows the ATWV improvements or degradation for queries of different length as the down-sampling rate is varied.
The systems with down-sampling generally perform better as the query length increases supporting the hypothesis that without down-sampling, the system struggles to model long-term dependencies.
On the other hand, as the rate of down-sampling is increased, the model struggles with detecting shorter queries (see for instance the performance with down-sampling rate of 16), supporting the idea that loss of resolution eventually limits feasible amount of down-sampling.

\subsection{Impact of multilingual pretraining}
\label{sec:experiments:performance}
In this section, we experiment with methods to increase the effective training data by using speed perturbation~\cite{ko2015audio} and multilingual pretraining.

Table~\ref{tab:feature_comp_and_sp} shows the effect of speed perturbation on KWS performance.
As is common practice, we create two extra copies of the training data by perturbing the speaking rates by factors of $0.9$ and $1.1$ respectively.
The bottleneck feature (BNF) rows of the table replicate the best results reported in~\cite{yusuf21_interspeech}.
We observe that speed perturbation leads to significant improvements regardless of input feature, with much higher relative improvements on MFCC.
Therefore, in subsequent experiments, we use speed perturbation by default.

\begin{table}[t]
    \caption{Term weighted values on the various development and evaluation sets.
    ``+sp" denotes the use of speed-perturbation and ``+Pretrain" indicates multilingual pretraining.}
    \centering
    \begin{tabular}{llcc}
    \toprule
         Language & System & Dev MTWV & Eval ATWV \\
         \midrule
         Pashto & MFCC & 4.9 &	6.9 \\
         &+sp & 7.9 & 10.6 \\
         & +Pretrain & \textbf{9.4} & \textbf{11.4} \\
         \cmidrule{2-4}
         & BNF & 11.7 & 15.2 \\
         &+sp & 14.5 & 18.1 \\
         & +Pretrain & \textbf{16.3} & \textbf{20.3} \\
         \midrule
         Turkish & MFCC & 18.1 & 7.9 \\
         &+sp & 24.4 & 13.5 \\
         & +Pretrain & \textbf{31.6} & \textbf{18.2} \\
         \cmidrule{2-4}
         & BNF &29.1& 17.8\\
         &+sp &31.0& 23.7\\
         & +Pretrain & \textbf{37.6} & \textbf{24.7} \\
         \midrule 
         Zulu & MFCC & 4.7 & 6.6 \\
         &+sp & 9.9 & 12.3 \\
         & +Pretrain & \textbf{15.9} & \textbf{17.7} \\
         \cmidrule{2-4}
         & BNF & 16.4 & 17.9 \\
         &+sp & 23.5 & 25.3 \\
         & +Pretrain & \textbf{26.4} & \textbf{26.8} \\
         \bottomrule
    \end{tabular}
    \label{tab:feature_comp_and_sp}
\end{table}

Table~\ref{tab:feature_comp_and_sp} also shows the effect of multilingual pretraining on KWS performance.
Here, we only use speed perturbation when finetuning, and not when pretraining in order to limit computational costs.
We observe significant improvements on the baseline, on top of the improvements from speed perturbation, regardless of input feature type.
This is despite the fact that the BNF already contain multilingual information.

\subsection{Effect of the ratio of negative to positive training utterances}
\label{sec:experiments:negatives}
In Section~\ref{sec:methods:training}, for each training phrase, we sample {${M}$} utterances, one of which is the utterance from which the current training phrase is taken.
The other {${M}-1$} ``negative" utterances are sampled randomly. In the experiments so far, we have set ${M}=4$.

\begin{figure}[t]
    \centering
    \resizebox{0.25\textwidth}{!}{%
        \begin{tikzpicture}
            \input{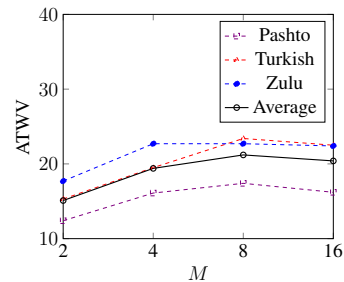}
        \end{tikzpicture}
    }
    \caption{ATWV on the evaluation sets as the ratio of negative training utterances is varied.}
    \label{fig:negatives}
\end{figure}

Figure~\ref{fig:negatives} shows the result of varying ${M}$ with BNF-based models.
Increasing ${M}$ has two obvious effects on the optimization; it reduces the approximation error due to the sampling process and it inadvertently up-weights the contribution of negative samples to the loss function (effectively reduces $\lambda$).
We argue that the ATWV improvements we observe are a result of the former, since, as we have already seen in Section~\ref{sec:experiments:loss} (and Figure~\ref{fig:loss_weight}), decreasing $\lambda$ does not improve the ATWV and doubling (or even quadrupling it) does not degrade the term weighted value to the extent that setting ${M}=1$ does.
However, the latter has an impact on STWV, which decreases strictly as ${M}$ increases although we do not report it here to reduce clutter.
This is due to a ``broken-clock" effect, where models trained with lower ${M}$ (similar to models trained with smaller $\lambda$) have higher recall simply by virtue of returning far more hits, whether spurious or correct, as a consequence of seeing a lower number and diversity of negative training utterances.
Finally, we see that with ${M}=8$, we get a good enough approximation and that increasing $M$ further does not lead to significant improvements in term weighted value--even slightly degrading the performance for Pashto.

Table~\ref{tab:twv_npd} shows the results of increasing ${M}$ from 4 to 8 for the speed-pertubed BNF models with and without pretraining.
We observe that increasing ${M}$ generally improves the performance, with the only exception being the Turkish Dev MTWV with pretraining and the Zulu Eval ATWV without pretraining.
We reiterate here that the variation in ${M}$ is only done for the finetuning. Pretraining is always done with ${M}=4$. We do not experiment with higher ${M}$ for pretraining due to the computational costs that would be involved.

\begin{table}[t]
    \caption{Term weighted value as the number of utterances per training step is varied with or without multilingual pretraining.}
    \centering
    \begin{tabular}{llccc}
    \toprule
         Language & System & ${M}$ & Dev MTWV & Eval ATWV \\
         \midrule
         Pashto & BNF + sp & 4 & 14.5 & 18.1 \\
         & BNF + sp & 8 & \textbf{16.1} & \textbf{20.2} \\
         \cmidrule{2-5}
         & +Pretrain & 4 & 16.3 & 20.3 \\
         & +Pretrain & 8 & \textbf{17.5} & \textbf{20.7} \\
         \midrule
         Turkish & BNF + sp & 4 & 31.0& 23.7\\
         & BNF + sp & 8 & \textbf{32.6} & \textbf{24.8} \\
         \cmidrule{2-5}
         & +Pretrain & 4 & \textbf{37.6} & 24.7 \\
         & +Pretrain & 8 & 35.0 & \textbf{27.0} \\
         \midrule 
         Zulu & BNF + sp & 4 & 23.5 & 25.3 \\
         & BNF + sp & 8 & \textbf{24.1} & 25.3 \\
         \cmidrule{2-5}
         & +Pretrain & 4 & 26.4 & 26.8 \\
         & +Pretrain & 8 & \textbf{27.7} & \textbf{27.2} \\
         \bottomrule
    \end{tabular}
    \label{tab:twv_npd}
\end{table}

\subsection{Performance of queries with various properties}
\label{sec:experiments:properties}
\begin{table*}[t]
    \centering
    \caption{Evaluation set IV and OOV term weighted values for the proposed system, a hybrid ASR-based KWS system and the fusion of both systems.}
    \begin{tabular}{llccccccccccccc}
        \toprule
         Metric & System  & \multicolumn{2}{c}{Pashto} && \multicolumn{2}{c}{Turkish} && \multicolumn{2}{c}{Zulu} && \multicolumn{2}{c}{Average}\\
         && IV & OOV && IV & OOV && IV & OOV && IV & OOV \\
         \midrule
         \multirow{3}{*}{ATWV} & ASR-based & 35.0 & 12.7 && 46.2 & 27.9 && 37.6 & 23.5 && 39.6 & 21.4 \\
         & Proposed & 22.0 & 13.5 && 29.0 & 26.7 && 27.9 & 25.3 && 26.3 & 21.8 \\
         & ASR-based + Proposed & \textbf{38.0} & \textbf{18.9} && \textbf{50.3} & \textbf{35.0} && \textbf{41.3} & \textbf{33.2} && \textbf{43.2} & \textbf{29.0}\\
         \midrule
         \multirow{3}{*}{OTWV} & ASR-based & 51.5 & 24.4 && 62.5 & 41.9 && 50.8 & 36.5 && 54.9 & 34.3 \\
         & Proposed & 39.3 & 28.4 && 46.1 & 45.4 && 41.3 & 39.7 && 42.2 & 37.8 \\
         & ASR-based + Proposed & \textbf{55.1} & \textbf{36.2} && \textbf{65.9} & \textbf{54.0} && \textbf{56.3} & \textbf{48.8} && \textbf{59.1} & \textbf{46.3} \\
         \bottomrule
    \end{tabular}
    \label{tab:comparison_with_asr_based}
\end{table*}

In this section, we analyze the performance of the proposed model on queries of various properties.
Specifically, we analyze how the performance of the model changes depending on whether the query in question is in-vocabulary or out-of-vocabulary, as well as how the performance changes with length of the query.
We compare the result to how the same factors affect conventional ASR-based systems.

For this comparison, we build a baseline KWS system based on a TDNN-based~\cite{peddinti2015time} hybrid ASR model.
To have a fair comparison, the TDNN acoustic model is pretrained on the same data we use for multilingual pretraining of the proposed model, and finetuned on each target language with speed perturbation applied for both pretraining and finetuning.
We use a word-subword hybrid index where IV queries are searched in a word-based index while OOV queries are searched in a syllabic one.
Both the word and the subword lattices are obtained using respective word and syllabic trigram Kneser-Ney-smoothed~\cite{ney1994structuring} language models which we found to perform better than higher order language models in this low-resource setting.
We use the official lexicon to get IV word pronunciations and to train a Sequitur grapheme-to-phoneme converter~\cite{bisani2008joint} for getting OOV pronunciations.

\subsubsection{IV vs OOV queries}
Table~\ref{tab:comparison_with_asr_based} shows the performance of the ASR-based KWS system and proposed system on IV and OOV queries.
We find that the proposed model has much smaller discrepancy between IV and OOV performance than the ASR-based system.
With the exception of Pashto, the ATWV difference between IV and OOV queries is always under 3 points.

In terms of ATWV, the proposed system slightly outperforms the ASR-based baseline system for OOV queries but lags it significantly for IV queries.
This is somewhat expected, especially in the low-resource settings, as the baseline has a strong guide for IV queries in the word language model, an advantage that is diminished for OOV queries, even with a subword language model and index.

We also report the OTWV, which slightly favors the proposed model compared to the baseline.
For IV terms, the relative average degradation between the proposed system and the baseline is reduced from 34\% in ATWV to 23\% OTWV.
For OOV terms, the relative improvement is increased from 2\% to 10\%.
This suggests that although the overall trends stay the same, part of the difference in performance is due to the score normalization being better suited to the baseline rather than qualitative differences in the model.

Finally, we report the results of fusion, where we combine the hitlists from both systems by weighted summation of scores with weights tuned on the development sets.
We find that the performance of the baseline is significantly improved by score fusion; around 9\% on IV and 36\% on OOV ATWV, with similar improvements in OTWV.
This underscores the potential benefit of deploying both systems in tandem where computationally feasible.

\subsubsection{Query length}
\begin{figure}[t]
    \centering
    \resizebox{0.275\textwidth}{!}{%
        \begin{tikzpicture}
            \begin{axis}[
    xlabel={Query length in letters},
    ylabel={ATWV difference from ASR-based},
    legend pos=north west,
]

    \addplot[color=black, mark=square,]
    coordinates {
        (2, 0.0)
(3, 0.0)
(4, 0.0)
(5, 0.0)
(6, 0.0)
(7, 0.0)
(8, 0.0)
(9, 0.0)
(10, 0.0)
(11, 0.0)
(12, 0.0)
(13, 0.0)
(14, 0.0)
(15, 0.0)
(16, 0.0)
(17, 0.0)
(18, 0.0)
(19, 0.0)
(20, 0.0)
(21, 0.0)
(22, 0.0)
    };
\addlegendentry{ASR-based}
    \addplot[color=red, mark=triangle,]
    coordinates {
        (2, 0.4366666666666667)
(3, -7.280000000000001)
(4, -4.046666666666667)
(5, -9.986666666666666)
(6, -10.593333333333334)
(7, -11.670000000000002)
(8, -15.00333333333333)
(9, -16.42)
(10, -15.456666666666672)
(11, -6.49666666666666)
(12, -4.630000000000005)
(13, -12.143333333333336)
(14, 1.6899999999999977)
(15, -6.459999999999994)
(16, 7.044999999999995)
(17, 9.445)
(18, 8.670000000000002)
(19, 16.009999999999998)
(20, 7.880000000000003)
(21, 20.63)
(22, 8.020000000000003)
    };
    \addlegendentry{Proposed}

    \addplot[color=blue, mark=*,]
    coordinates {
        (2, 0.0)
(3, 2.2533333333333325)
(4, 3.7999999999999994)
(5, 2.530000000000001)
(6, 4.686666666666667)
(7, 4.876666666666665)
(8, 2.7933333333333317)
(9, 4.056666666666662)
(10, 1.98666666666666)
(11, 3.373333333333335)
(12, 7.046666666666671)
(13, 10.880000000000004)
(14, 8.370000000000005)
(15, 12.579999999999998)
(16, 19.654999999999994)
(17, 19.235)
(18, 11.339999999999996)
(19, 22.22)
(20, 18.060000000000002)
(21, 42.06)
(22, 3.080000000000002)
    };
    \addlegendentry{ASR-based + Proposed}
\end{axis}
        \end{tikzpicture}
    }
    \caption{Average difference in ATWV of various systems when compared to the ASR-based baseline as query length varies.
    }
    \label{fig:atwv_kwlen_dist}
\end{figure}

\begin{figure}[t]
    \centering
    \resizebox{0.275\textwidth}{!}{%
        \begin{tikzpicture}
            \begin{axis}[
    xlabel={Query length in letters},
    ylabel={OTWV difference from ASR-based},
    legend pos=north west,
]

    \addplot[color=black, mark=square,]
    coordinates {
        (2, 0.0)
(3, 0.0)
(4, 0.0)
(5, 0.0)
(6, 0.0)
(7, 0.0)
(8, 0.0)
(9, 0.0)
(10, 0.0)
(11, 0.0)
(12, 0.0)
(13, 0.0)
(14, 0.0)
(15, 0.0)
(16, 0.0)
(17, 0.0)
(18, 0.0)
(19, 0.0)
(20, 0.0)
(21, 0.0)
(22, 0.0)
    };
\addlegendentry{ASR-based}
    \addplot[color=red, mark=triangle,]
    coordinates {
        (2, 1.1766666666666665)
(3, -4.576666666666668)
(4, -5.583333333333333)
(5, -11.696666666666667)
(6, -11.926666666666668)
(7, -9.51)
(8, -14.136666666666656)
(9, -11.74333333333333)
(10, -10.813333333333333)
(11, -2.646666666666666)
(12, 0.2366666666666788)
(13, -7.453333333333347)
(14, -0.3100000000000023)
(15, -6.2050000000000125)
(16, 6.589999999999996)
(17, 12.325000000000003)
(18, 7.329999999999998)
(19, 17.650000000000002)
(20, 11.110000000000007)
(21, 7.940000000000005)
(22, 9.25)
    };
    \addlegendentry{Proposed}

    \addplot[color=blue, mark=*,]
    coordinates {
        (2, 1.1766666666666665)
(3, 9.89333333333333)
(4, 6.61666666666666)
(5, 3.3433333333333337)
(6, 3.816666666666663)
(7, 5.786666666666662)
(8, 2.6533333333333453)
(9, 6.106666666666674)
(10, 3.3433333333333337)
(11, 8.333333333333343)
(12, 14.810000000000002)
(13, 12.049999999999992)
(14, 11.229999999999997)
(15, 13.934999999999988)
(16, 19.665)
(17, 19.36999999999999)
(18, 13.770000000000003)
(19, 23.529999999999998)
(20, 19.450000000000003)
(21, 28.580000000000005)
(22, 14.810000000000002)
    };
    \addlegendentry{ASR-based + Proposed}
\end{axis}
        \end{tikzpicture}
    }
    \caption{Average difference in ATWV of various systems when compared to the ASR-based baseline as query length varies.}
    \label{fig:otwv_kwlen_dist}
\end{figure}
We have seen that while our approach does not distinguish much between IV and OOV queries, its IV performance trails that of a strong ASR-based KWS baseline.
To find the root of this difference, we compare the performance of the systems on queries of various length.

Figures~\ref{fig:atwv_kwlen_dist}~and~\ref{fig:otwv_kwlen_dist} show the average (across languages) difference in ATWV and OTWV respectively between the proposed model and the baseline.
Negative values indicate query lengths for which the baseline is better and positive values indicate query lengths for which the proposed system is better.
Note that although it is not conveyed in this figures, all systems---including the baseline---have better performance as the query length increases.
Even so, we find that in general, the baseline performs better for shorter queries, both systems perform comparably for mid-length queries, and the proposed system performs better for long queries (above 15 characters).
Finally, we observe that fusion outperforms the baseline regardless of query length.

\begin{figure}[t]
    \centering
    \resizebox{0.25\textwidth}{!}{%
        \begin{tikzpicture}
            \begin{axis}[
    xlabel={Query length in letters},
    ylabel={CMF},
    legend pos=south east,
    legend style={nodes={scale=1.4, transform shape}}
]
    \addplot[color=black, mark=square, thick]
    coordinates {
        (2, 0.012256973795435333)
(3, 0.12341504649196956)
(4, 0.31234150464919697)
(5, 0.5194420963651732)
(6, 0.687235841081995)
(7, 0.8284023668639053)
(8, 0.9087066779374472)
(9, 0.9560439560439561)
(10, 0.981403212172443)
(11, 0.9928148774302621)
(12, 0.9970414201183432)
(13, 1.0)
    };
    \addlegendentry{Pashto}

    \addplot[color=red, mark=triangle, thick]
    coordinates {
        (2, 0.0)
(3, 0.014232673267326733)
(4, 0.09467821782178218)
(5, 0.2691831683168317)
(6, 0.4443069306930693)
(7, 0.5909653465346535)
(8, 0.7085396039603961)
(9, 0.8007425742574258)
(10, 0.8688118811881188)
(11, 0.9170792079207921)
(12, 0.9517326732673267)
(13, 0.9721534653465347)
(14, 0.989480198019802)
(15, 0.9956683168316832)
(16, 1.0)
    };
    \addlegendentry{Turkish}

    \addplot[color=blue, mark=*, thick]
    coordinates {
        (2, 0.0014265335235378032)
(3, 0.007845934379457917)
(4, 0.02781740370898716)
(5, 0.09415121255349501)
(6, 0.19258202567760344)
(7, 0.3145506419400856)
(8, 0.39229671897289586)
(9, 0.46790299572039945)
(10, 0.5499286733238231)
(11, 0.6255349500713266)
(12, 0.6997146932952925)
(13, 0.7738944365192582)
(14, 0.833095577746077)
(15, 0.8880171184022825)
(16, 0.920827389443652)
(17, 0.9500713266761769)
(18, 0.9679029957203994)
(19, 0.9800285306704708)
(20, 0.9885877318116976)
(21, 0.9935805991440799)
(22, 1.0)
    };
    \addlegendentry{Zulu}

\end{axis}
        \end{tikzpicture}
    }
    \caption{Cumulative distributions of query lengths for each evaluation set.}
    \label{fig:kwlen_dist}
\end{figure}
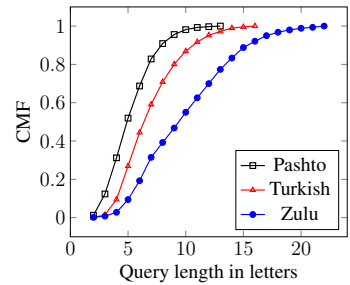

These results are not surprising as short queries are both easier to miss and easier to falsely spot, and the ASR-based system being able to leverage contextual information provided by the language model helps it better find short queries.
They also explain why our model performs significantly worse for Pashto than the other two languages.
As Figure~\ref{fig:kwlen_dist} shows, the queries in the Pashto evaluation set are quite short, with over 50\% of queries being less than 5 letters long.

\section{Conclusion}
\label{sec:conclusion}
In this paper, we extend our recent work~\cite{yusuf21_interspeech} on end-to-end keyword search.
Our model provides a simplified pipeline for keyword search, comprising a pair of encoders: one for encoding spoken archives, and a second for text queries.
Keyword search is then effected in the resulting vector-spaces by computing inner-products between the document and query encodings.
Compared to~\cite{yusuf21_interspeech}, in this work, we explore multilingual pretraining and conduct thorough analyses of various components, strengths and weaknesses of the proposed model.
Our experiments show that:
\begin{itemize}
    \item Our model significantly benefits from multilingual pretraining, with considerable increase in term weighted values without making the model more complex.
    \item Our model retrieves out-of-vocabulary queries almost as well as it retrieves in-vocabulary ones, and slightly outperforms a strong ASR-based keyword search system on OOV queries.
    \item The simplicity of our model comes at the cost of worse performance than the ASR-based system on IV queries and short queries---two query types where the ASR-based system benefits from contextual clues provided by the language model.
    \item Our approach is complementary with the ASR-based system, and combining the two improves the performance of the ASR-based system, even for IV queries and short queries.
\end{itemize}

Our model has two main limitations, which provide avenues for future work:
\begin{itemize}
    \item The model does not use linguistic context information, making it worse that the ASR-based system on IV queries and short queries.
    It would therefore be worth exploring methods to incorporate external text without complicating the inference, similar to joint text and speech training for end-to-end ASR~\cite{wang21t_interspeech,yusuf22usted,thomas2022integrating}.
    \item The document representation grows linearly with the size of the archive.
    Although inner-products can be efficiently computed even for very large indices, storing those indices in memory becomes untenable for archives larger than a few hundred hours.
    Potential solutions include quantization techniques such as binary hashing~\cite{jansen2011efficient} and product-quantization~\cite{jegou2010product} to reduce both the storage and search computation costs.
\end{itemize}

\section{Acknowledgments}
\label{sec:acknowledgment}
This work was supported by European Union’s Horizon 2020 project No. 870930 - WELCOME, the Czech National Science Foundation (GACR) project ``NEUREM3" No. 19-26934X,
the Turkish Directorate of Strategy and Budget under the TAM Project (2007K12-873) and the ROYAL Project (2019K12-149250) as well as the Boğaziçi University Research Fund under the Grant Number 16903.
Computing on IT4I supercomputer was supported by the Czech Ministry of Education, Youth and Sports through the e-INFRA CZ (ID:90140).

\bibliographystyle{IEEEtran}
\bibliography{my_bib}

\end{document}